\def\year{2020}\relax
\documentclass[letterpaper]{article} 
\usepackage{aaai20}  
\usepackage{times}  
\usepackage{helvet} 
\usepackage{courier}  
\usepackage[hyphens]{url}  
\usepackage{graphicx} 
\usepackage{graphicx}  
\title{Emergence of Writing Systems Through Multi-Agent Cooperation}
\author{\Large \textbf{Shresth Verma, Joydip Dhar}\\ 
ABV-Indian Institute of Information Technology and Management\\ 
Gwalior, MP, India 474003\\
E-mail: vermashresth@gmail.com, jdhar.iiitmg@gmail.com 
}
 
\begin{document}

\maketitle

\begin{abstract}
Learning to communicate is considered an essential task to develop a general AI. While recent literature in language evolution has studied emergent language through discrete or continuous message symbols, there has been little work in the emergence of writing systems in artificial agents. In this paper, we present a referential game setup with two agents, where the mode of communication is a written language system that emerges during the play. We show that the agents can learn to coordinate successfully using this mode of communication. Further, we study how the game rules affect the writing system taxonomy by proposing a consistency metric. 
\end{abstract}

\noindent

\section{Introduction}
Recent advances in deep learning have shown exceptional results in language-related tasks such as machine translation, question answering, or sentiment analysis. However, the supervised approaches that capture the underlying statistical patterns in language are not sufficient in perceiving the interactive nature of communication and how humans use it for coordination. It is thus crucial to learn to communicate by interaction, i.e., communication must emerge out of necessity.
Such study gives further insights into how communication protocols emerge for successful coordination and the ability of a learner to understand the emerged language.

Several recent works \cite{lazaridou2016multi,havrylov2017emergence,lazaridou2018emergence,mordatch2018emergence}, have shown that in multi-agent cooperative setting of referential games, deep reinforcement learning can successfully induce communication protocols. In these games, communication success is the only supervision during learning, and the meaning of the emergent messages gets grounded during the game.
In \cite{lazaridou2016multi}, the authors have restricted the message to be a single symbol token picked from a fixed vocabulary while in \cite{havrylov2017emergence}, the message is considered to be a sequence of symbols. \cite{lazaridou2018emergence} demonstrates that successful communication can also emerge in environments which present raw pixel input. \cite{mordatch2018emergence} further extends the scope of mode of communication by also studying the emergence of non-verbal communication. 

While these works have studied a wide variety of game setups as well as variations in communication rules, none of them have considered written language system as a mode of communication. Historically, written language systems have shown complex patterns in evolution over time. Moreover, the process of writing requires sophisticated graphomotor skills which involves both linguistic and non-linguistic factors. Thus writing systems can be considered crucial for understanding autonomous system development.
We are further motivated by the work in \cite{ganin2018synthesizing}, where the authors demonstrate that artificial agents can produce visual representations similar to those created by humans.
This can only be achieved by giving them access to the same tools that we use to recreate the world around us. 
We extend this idea to study emergence of writing systems.

\section{Referential Game Framework}
In our work, we have used two referential game setups that are slight modifications to the ones used in \cite{lazaridou2016multi,lazaridou2018emergence}.

There are two players, a sender and a receiver. From a given set of images $I = \{{i_j}\}_{j=1}^N$, we sample a target image $t \in I$ and $K - 1$ distracting images $D =\{{d_j}\}_{j=1}^{K-1}, d_j \in I$ s.t. $\forall j\: t \neq d_j$. Now, we define two sender types, \textit{Distractor Agnostic (D-Agnostic)}: where the sender only has access to the target image $t$;  \textit{Distractor Aware (D-Aware)}: where the sender has access to the candidate set $ C = t \cup D$.
In both these variations, the sender has to come up with a message $M_l = \{m_j\}_{j=1}^{l}$, which is a sequence of $l$ brushstrokes. A black-box renderer $\mathcal{R}$ accepts the sequence of brushstrokes $M_l$ and paints them onto a canvas. This results in a written symbol image $W = \mathcal{R}(M_l)$.
Given the written symbol image $W$ and the candidate set C, the receiver has to identify the target image $t$. Communicative success is achieved when the target is correctly identified and a payoff of 1 is assigned to both the players. In rest of the cases, payoff is 0.

    

\section{Experimental Setup}

\subsection{Agents}
The sender and receiver are modelled as reinforcement learning policy networks $S_\theta$ and$R_\phi$. Specifically, the sender is a recurrent neural network which takes as input the current state of the canvas along with the visual input $V$ which can either be target image $t$ (D-Agnostic) or candidate set $C$ (D-Aware). At the $i^{th}$ timestep, the sender outputs a brushstroke $m_i$. The canvas state is the intermediate rendering $\mathcal{R}(M_i)$,  where $M_i$ is the collection of brushstrokes produced upto timestep $i$. Thus, $m_{i+1}$ is generated by sampling from $ S_{\theta}(\mathcal{R}(M_i), h_i, V)$ where $h_i$ is the internal hidden state maintained across timesteps. The sequence is terminated when either the maximum sequence length $L$ is reached or a terminal flag is produced along with the brushstroke.
The internal state is maintained across timesteps using an LSTM cell \cite{hochreiter1997long}.
The receiver agent first extracts features from the written symbol image $W$. For creating brushstrokes that are similar to written languages used by humans, we use feature extractor from a Siamese Neural Network \cite{koch2015siamese}, pre-trained on the OMNIGLOT dataset \cite{lake2015human}. Given the written symbol image $W$, a candidate set U (a random permutation of C), and the feature extractor $f_s$, the receiver returns an integer value $t' = R_{\phi}(f_s(W), U)$ in the range 0 to K-1 that points to the target.

\subsection{Learning}
For both the agents, we pose the learning of communication protocols as maximization of the expected return $E_{\tilde{r}}[R(\tilde{r})]$, where $R$ is the reward function.
The payoff is 1 for both the agents iff $R_{\phi}(f_s(S_{\theta}(\mathcal{R}(M_i), h_i, V)), U) = t$ , where $i$ is the last timestep of the episode. In all other cases and intermediate timesteps, the payoff is 0.
Because of the high dimensional search space introduced due to brushstrokes, we use Proximal Policy Optimization (PPO) \cite{schulman2017proximal} for optimizing the weights of sender and receiver agents.

\subsection{Images}
We have used CIFAR-10 dataset \cite{krizhevsky2009learning}, as a source of images. From the test set of CIFAR-10, we randomly sample 100 images from each class and represent them as outputs from $relu7$ layer of pre-trained VGG-16 convNet \cite{simonyan2014very}.

\section{Results and Conclusion}
\begin{figure}
    \centering
    \includegraphics[width=.7\linewidth]{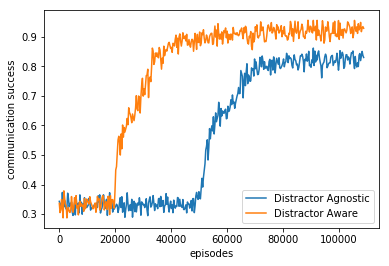}
    \caption{Communication success as a function of training episodes for referential games with K = 3 and L = 2}
    \label{fig:my_label}
\end{figure}

Figure \ref{fig:my_label} shows the performance of our game setup for both the sender variations. The agents converge to coordination in both sender types, but D-Aware sender reaches higher levels more quickly.
Further, we quantify the consistency of a writing system by studying the variability of the symbols produced for a given entity $e$. Let $w_e$ be the set of all written symbol images representing $e$. We define heatmap $H_e = mean(w_e)$. For a writing system consistent for the entity $e$, $H_e$ would contain sharp brushstrokes while a non-consistent writing system would give a blurred heatmap. We thus compute Variance of Laplacian (VoL) of the heatmap to quantify sharpness. Table 1 reports the average consistency score given by \[\frac{\sum_{e \in E}VoL(H_e)}{|E|}\] where $E$ is the set of all the entities considered which can either be targets ($t$) or target-distractor combinations ($t\&d$).
We also report a baseline consistency score where heatmap is generated by averaging across the universal set of generated symbol images.

High consistency of D-Agnostic sender indicates a one-to-one mapping from target class to written symbols. The D-Aware sender has low consistency over target class but high consistency for target-distractor combinations . This means that symbols are context dependent. From our qualitative evaluations, we infer that D-Aware sender assigns meaning to brushstrokes that represent conceptual differences between target and distractors. Furthermore, D-Agnostic sender uses a scheme akin to hierarchical encoding to attribute high level semantics to brushstrokes. Thus, the writing system emerging from D-Aware sender is an ideographic one representing concepts while D-Agnostic sender produces a writing system which has compositionality and shows logographic traits.

\begin{table}[]
\centering
\footnotesize
\begin{tabular}{|l|p{20mm}|p{25mm}|}
\hline
\textbf{Sender Type} & \textbf{Avg. Consistency Score} & \textbf{Baseline Consistency Score} \\ \hline
D-Agnostic$^{t}$  & 0.019                      & 0.0055                                  \\ \hline
D-Aware$^{t}$     & 0.007                     & 0.0044                                   \\ \hline
D-Aware$^{t\&d}$     & 0.015                      & 0.0044                                  \\ \hline
\end{tabular}
\caption{Consistency Score for different sender types}
\label{lab:cons}
\end{table}
\bibliography{bibi.bib}
\bibliographystyle{aaai}

\end{document}